\documentclass[twocolumn,aps]{revtex4}

\usepackage{dcolumn}
\usepackage{graphicx}

\newcommand{\be}{\begin{equation}} 
\newcommand{\ee}{\end{equation}} 
\newcommand{\bea}{\begin{eqnarray}} 
\newcommand{\eea}{\end{eqnarray}} 
\newcommand{\bc}{\begin{center}} 
\newcommand{\ec}{\end{center}}

\newcommand{\yarim}{{1\over 2}}

\begin{document}
\bibliographystyle{plain}
\pagenumbering{arabic} 

\title{Strategies for the Evolution of Sex}

\author{Erkan T{\"u}zel$^{1,2}$, Volkan Sevim$^{1}$ and Ay{\c s}e 
Erzan$^{1,3}$}

\affiliation{$^1$ Department of Physics, Faculty of Sciences and Letters\\
Istanbul Technical University, Maslak 80626, Istanbul, Turkey \\
$^2$ Department of Physics, Faculty of Sciences and Letters \\
        I{\c s}{\i }k University, Maslak, 80670, Istanbul, Turkey\\
$^3$ G\"ursey Institute, P.O.B. 6, \c Cengelk\"oy, 81220 Istanbul, Turkey
}
\date{\today}

\begin{abstract}

We find that the hypothesis made by Jan, Stauffer and Moseley [Theory in 
Biosc., 
{\bf 119}, 166 (2000)]
for the evolution of sex, namely a strategy
devised to escape extinction due to too many deleterious mutations, is   
sufficient but not necessary for the successful evolution of a steady    
state population of sexual individuals within a finite population. Simply
allowing for a finite probability for conversion to sex in each generation 
also 
gives rise to
a stable sexual population, in the presence of an upper limit on the number of 
deleterious mutations per individual. For large values of this probability, we 
find 
a  phase transition to an intermittent, multi-stable regime.  On the other 
hand, 
in the limit of extremely slow drive, another transition takes place to a 
different 
steady state distribution, with fewer deleterious mutations within the asexual 
population.


\end{abstract}

\maketitle

\section{Introduction} 

In a previous paper~\cite{Orcal} we have shown that a finite steady state
sexual population may arise from a purely asexual one, if an excess of   
deleterious mutations causes the individual to engage in sex as a means of
escaping death~\cite{Jan}.  Since sexual reproduction led, in our model, to
diploidy, it implied also the adoption of a mechanism for preferential
expression of certain genes, and we assumed deleterious mutations to be
recessive.~\cite{Stauffer} 
 Under various different assumptions regarding the subsequent mode of
reproduction (i.e., whether sexual reproduction is hereditary or not) and
of the  number of offspring, we found that the diploid population always
persisted, and that it was consistently more successful in escaping the
effects of deleterious mutations. 

In the case where sexual reproduction was 
only practiced as a means of escaping death from too many deleterious 
mutations, but 
diploid cells were also allowed to multiply by mitosis, 
diploid individuals completely took over the population. Thus, in one of
our models (Model I) \cite{Orcal}, we were able to demonstrate a possible 
scenario for
the evolution of the analogue of a `haploid - diploid cycle'' ~\cite{Smith} 
where the 
organisms produce by asexual reproduction as long as they are reasonably fit 
(or the conditions 
reasonably favorable) but engage in sexual reproduction when the going gets 
tough.~\cite{Candan1,Candan2}

In this paper we will test whether a threshold mechanism for switching to 
sexual 
reproduction  is {\em
necessary} for the successful establishment of a sexual population.
We simulate two strategies for the evolution of sex within a fixed
population $N$ of simple organisms, who are all initially asexual (and
haploid), and subject to a constant rate $\Gamma $ of random mutations.  
Both haploid and diploid organisms die when the number of their expressed 
deleterious mutations exceed a certain number. (Henceforth we 
will omit ``deleterious" where it is evident that we mean a change away from 
the 
wild type gene.)

The first strategy (Model $A$) is the adoption, with a certain probability, of 
sexual reproduction and
consequent diploidy when the number of mutations exceeds a
threshold, threatening extinction. 
The second strategy (Model $B$) involves a small but constant probability
$\sigma$ for the accidental conversion to sex, independently of the number
of mutations (or, equivalently, the fitness) of the individual.
The cloning of sexual individuals is not allowed in either Model $A$ or $B$. 
On the other hand, we have tested for
 the effect of hereditary (habitual) v.s. non-hereditary (non-habitual) sex. 

In the implementation of both models, we have adopted a more realistic set
of rules for the mechanism of dominance, that is, the expression of mutated 
alleles,  than in our previous paper~\cite{Orcal}. 
Here we allow  a mutated gene to be expressed if  the cell is homozygous for 
mutated alleles at this locus. Hence, the number of expressed deleterious 
mutations for diploid individuals is the number $m$ of different loci at which 
the cell is homozygous for mutated alleles.

We find that {\em both} strategies $A$ and $B$ lead
to a finite steady state sexual population, with typically a smaller
average number of mutations (greater fitness) than the asexual population.  
Thus 
no   
threshold mechanism seems to be necessary for a successful sexual   
population  to take hold. However, for habitual practice of sexual 
reproduction by 
diploid individuals (i.e., those that are not facing extinction in Model $A$) 
calls for 
unrealistically large mutation rates in order for a macroscopic sexual 
population to 
be established in the steady state.

The organisation of the paper is as follows.  In the next section we explain 
in 
detail the two models and we report the result of our simulations.  In Section 
III, we display and examine the mean field evolution equations and discuss 
our findings in the light of these equations. In Section IV, we investigate 
the limits of 
strong and extremely weak driving of this system, for $\Gamma \to 1$ and 
$\Gamma \to 0$, as 
well as a transition to chaos via an intermittent route, found for 
large values of $\sigma$.  A discussion of the results from similar models and 
directions for 
further research are provided in section V.

\section{Models for conversion to sex}

We represent the genetic code of each one-celled
individual with a bit-string of ``0"s and ``1"s, after the
Penna model~\cite{Penna}. 
At each locus, we have taken the value ``0" to correspond to the wild type and 
``1," 
to a deleterious mutation (which we will call ``mutation," for short, where 
this is not liable to lead to any confusion.)
We use the bit
defining the ``sign", to specify whether the individual is 
asexual (+) or sexual (-). 
For asexual, haploid, cells, we have one 15-bit string,
whereas, for the sexual cells, we have two 15-bit
strings which are allowed to be different, i.e., the
individuals are now diploids.  

A mutation consists of flipping a randomly chosen bit except the sign bit, and 
it is 
implemented by scanning all the individuals in the population, and, with 
probability $\Gamma$ picking those to be mutated.
Clearly there may be any number of mutated individuals at any one generation 
(time step), the number fluctuating around $\Gamma N$,
where $N$ denotes the total population.

The number of deleterious mutations $m$ is simply the number of ``1"s for a 
haploid individual. For a diploid, the number of ``expressed" deleterious 
mutations is 
taken to be the number of loci at which both homologous alleles are set to 
``1."
This is how we model the mechanism of dominance of the wild type (or, 
equivalently, the 
recessiveness of deleterious mutations.)  We will use the term ``fitness" 
loosely, 
for $L-m$; thus increasing $m$ will decrease the fitness of the individual.

In the steady state, the distribution of the asexual and  sexual populations 
over 
 $m$,
are independent of $\Gamma$, for $\Gamma > 1/N$.
The cases where
$\Gamma < {1\over N}$ and $\Gamma \approx 1$ have interesting
consequences, and are discussed in section IV.     

The probability of survival as a function of $m$ 
 is given by a Fermi-like
distribution~\cite{Thoms},
$P(m)$,
\be P(m) = { 1  \over {\exp[\beta (m - \mu)] + 1}} \;\;\;.
\label{Fermi} \ee
For large $\beta$ (or ``low temperatures," in the language of
statistical mechanics), $P(m)$ behaves like a step function~\cite{Orcal}.
Individuals with $m>\mu$ die, those with $m<\mu$ survive, and those
with $m=\mu$ survive with a probability of $1/2$. In the simulations
we set $\beta=10$ and $\mu=4$.

We keep the total population constant, as in the Redfield
model~\cite{Redfield}, by making up for the deficit in the population
after all the bacteria have been either found fit for survival or
killed off according to the survival probability in
Eq.~(\ref{Fermi}). Asexual individuals multiply by simply making
another copy of themselves, namely by mitosis, while a pair of sexual 
organisms each contribute one bit-string to their offspring and die in 
the process. 

We performed the simulations on a fixed population of $N=1000$, 
for 16-bit strings. The total number of time steps in each simulation 
is taken to be much larger than the time necessary for  the transients to die 
off 
and the system to settle 
down to a steady state. Since the probability to mutate a single gene in a 
diploid
individual is  $\Gamma /(2L)$, on the average the steady state is reached 
after $2L/\Gamma$ time steps, where $2L$ is the total number of genes in a 
diploid individual, or, in other words, the number of mutated genes in the 
population 
 is greater 
than the total number of genes of one
individual. In all the simulations, the reported results are averages over 10 
runs. 
The fluctuations
depend on the model chosen, however the relative error estimate 
based on one standard deviation is typically less than about $6 \%$, as long 
as there 
is only one fixed point for the dynamics.  We will report on situations 
where we encounter an intermittent route to chaos in Section IV.

\subsection{Asexual steady state}

We start with a set of $N$ initially identical asexual 
individuals,
all identical to the {\em wildtype}, i.e., all $0$'s. Under the 
conditions outlined above, without introducing sex, the population
of asexuals settles down to the steady state 
distribution given in~\cite{Orcal}
for $\Gamma \geq {1 \over N}$ namely, 
$n_H(m)/N_A = 0.012,0.098,0.356,0.531,0.001$ for $m=0, \ldots,4$.
In this region this steady state distribution is independent of 
$\Gamma$, which only sets the scale of time. That this should be 
so, is not self-evident, and only follows from the form of the 
solution to the set of evolution equations, as shown in Section III.

\subsection{Triggering sex}

The alteration of the sex gene can be accomplished in two different ways. One 
can choose to trigger sex with a threshold mechanism or define a constant
probability for each individual to become sexual. These mechanisms are
further discussed in the following subsections.  In either case, the haploid 
organism first makes a copy of its own set of genes, as if it were going to 
perform mitosis, but then forms two gametes instead.  One of these gametes 
will pair up with a gamete from another individual who has been turned on to 
sex, 
and the other will be discarded.
One should note that sexual reproduction may be implemented in 
different ways, resulting in different numbers of offsprings 
produced~\cite{Orcal}.
In this paper, we will define sexual reproduction in such a way that 
 when two sexual individuals mate they always give rise 
to one sexual offspring; thus,  the population is reduced by one, each
time an act of sexual reproduction takes place. Clearly, increasing the number 
of offspring 
will increase the advantage that the sexual population enjoys.  Indeed, 
judging 
from our previous results~\cite{Orcal}, the number of offspring exceeding two 
would lead to the takeover of the 
population by the dipoid sexual types.

When two diploid cells engage in sexual reproduction, they each contribute one 
gamete towards a single diploid sexual offspring.
Let us denote the two gametes  as $\{A a\}$ in one parent, and $\{B b\}$ 
in the other parent. 
Then the genome of the offspring may be,  $\{AB\}$, $\{Ab\}$,
$\{aB\}$ or $\{ab\}$. 
We do not allow for crossover between the gametes during sexual reproduction.

\subsubsection{Sex at the threshold of extinction - Model $A$}

In model $A$, alteration of the sex gene takes place only under special 
conditions, namely the threat of death due to too many 
mutations~\cite{Jan}. Once the asexual steady state is reached, we 
allow the sex gene to be ``turned on'' for the least fit members of the 
population.
In any pass through the population, if those individuals that are in the tail 
of 
the distribution  (i.e. those with $m \geq \mu$ mutations) survive, then they 
are turned sexual by deterministically and irreversibly switching their sign
bits to one. Once their sex bit is turned on, these individuals will
be labelled ``sexually active'' and mate with other sexually active 
individuals. 
If there is
only one active sexual at a certain time step then it must wait subsequent 
generations 
until it finds
a partner.
 After mating, the sexual individual becomes sexually inactive and the 
only
way for it to become sexually active again is to face extinction once more. 
The deficit in the population due to deaths and to sexual reproduction is then 
made up by 
copying randomly selected asexual individuals. 


\begin{figure}[!ht]
\leavevmode
\rotatebox{270}{\scalebox{.4}{\includegraphics{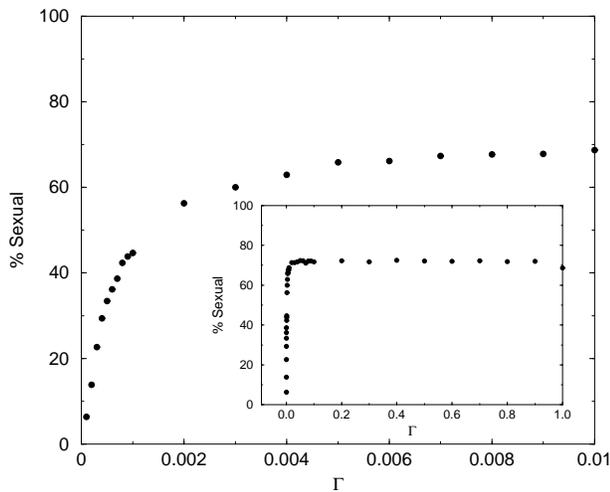}}}
\caption{The percentage of sexual population v.s. $\Gamma$ is plotted for 
Model $A$
where hereditary sex is not allowed, for a population of 1000 individuals.
The inset shows a larger range of  $\Gamma$ where the step-function like
jump is more apparent. Both curves represent averages over 10 runs.}
\end{figure} 

In this model, therefore, there is no hereditary sexuality: there is, however,
a hereditary transition to diploidy. This gives an unfair advantage to the
sexuals in that they both enjoy the benefits of diploidy and escape 
the disadvantage of $2 \rightarrow 1$ reproduction.

We see that for  $\Gamma \sim 10/N$ the proportion of the sexuals in the 
population 
saturates to  $\sim 70 \% $ as shown in Figure 1, and remains at this value 
independently 
of the  value of $\Gamma$. In order to obtain points 
near $\Gamma \simeq 0$ one has to do very long runs to get accurate results, 
and 
these are discussed in Section IV, as well as the chaotic behaviour displayed 
when $\Gamma$ 
becomes too close to 1.


\begin{figure}[!h]
\leavevmode
\rotatebox{270}{\scalebox{.36}{\includegraphics{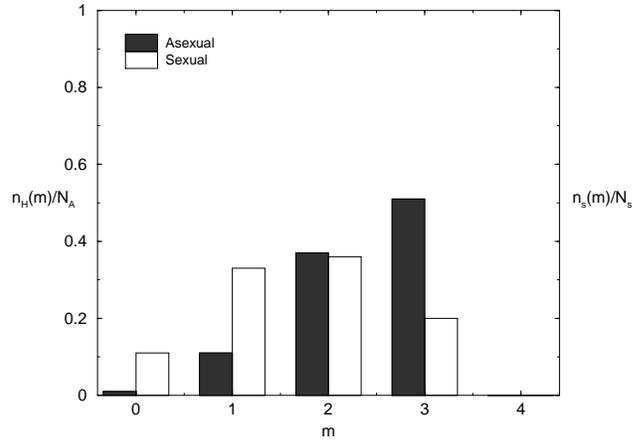}}}
\caption{The distribution of both sexuals and asexuals
over the number of expressed deleterious mutations $m$, 
for Model $A$. $\Gamma=10^{-3}$. Hereditary sexuality
is not allowed and the distributions are normalized to unity 
over each population separately.}
\end{figure} 


\begin{figure}[!ht]

\leavevmode
\rotatebox{270}{\scalebox{.4}{\includegraphics{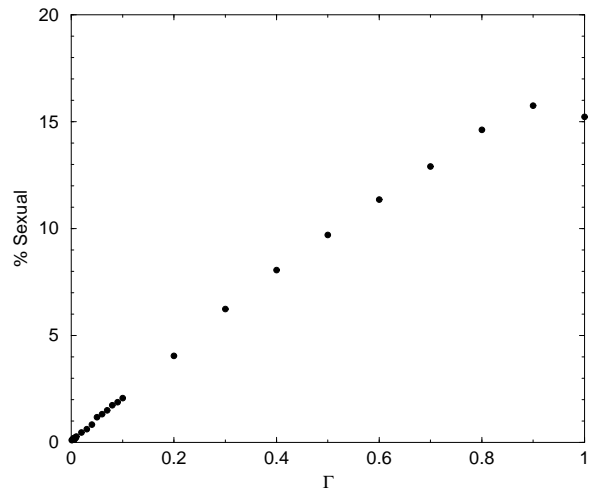}}}
\caption{The percentage of the sexual population v.s. $\Gamma$ for Model $A$ 
with
hereditary sexuality introduced. The total population is 1000 individuals
and the results are averaged over 10 runs.}
\end{figure} 

The steady state distributions of both asexuals and sexuals with respect to 
$m$ 
are also independent of $\Gamma$, 
 (See Figure 2) for $\Gamma \geq {1 \over N}$ and sufficiently smaller than 1.
The peak of the distribution shifts towards lower $m$ values for sexuals
as a result of the salutary effect of dominance in diploidy, and the 
reshuffling 
effect of sexual reproduction~\cite{Orcal}.

{\it Model $A$ with Hereditary Sex}

We have also tested the case of hereditary, or habitual, sex, 
in which sexually active individuals 
can mate randomly either with sexually active individuals who have been 
converted to sex in that generation, or with  individuals who have already 
been converted in some previous generation. As in the case of 
non-hereditary sex, the population is allowed to grow back to its fixed 
value by cloning randomly selected asexual units. 

This small difference results
in a noticeable increase in the number of matings at each time
step, and therefore leads to a decrease in the number of sexual individuals in 
the 
steady state. We have found that the steady state 
comprises a macroscopic sexual population only  for $\Gamma > 1/N$.  For 
$\Gamma < 1/N$, the average 
number of sexual individuals drops to about $1 \%$, or around 10 individuals 
in a population of $N=1000$.
The sexual population 
increases linearly with $\Gamma$ and reaches only $\sim 15 \% $ (as compared 
to 
$70 \%$ for non-hereditary sex) as  $\Gamma \simeq 1$ (see Fig. 3). 
The $m$-distributions are shown in Figs. (4a,b) for the asexual 
and sexual populations.  The peak of the sexual population has shifted to 1 as 
a result
of the greater number of mating events.
Thus we may conclude that hereditary and habitual sex in this model is 
punished 
more severely; the relative improvement in the mean value of $m$ does not 
compensate sufficiently 
for the loss of the parents.


\begin{figure}[!ht]
\leavevmode
\rotatebox{270}{\scalebox{.4}{\includegraphics{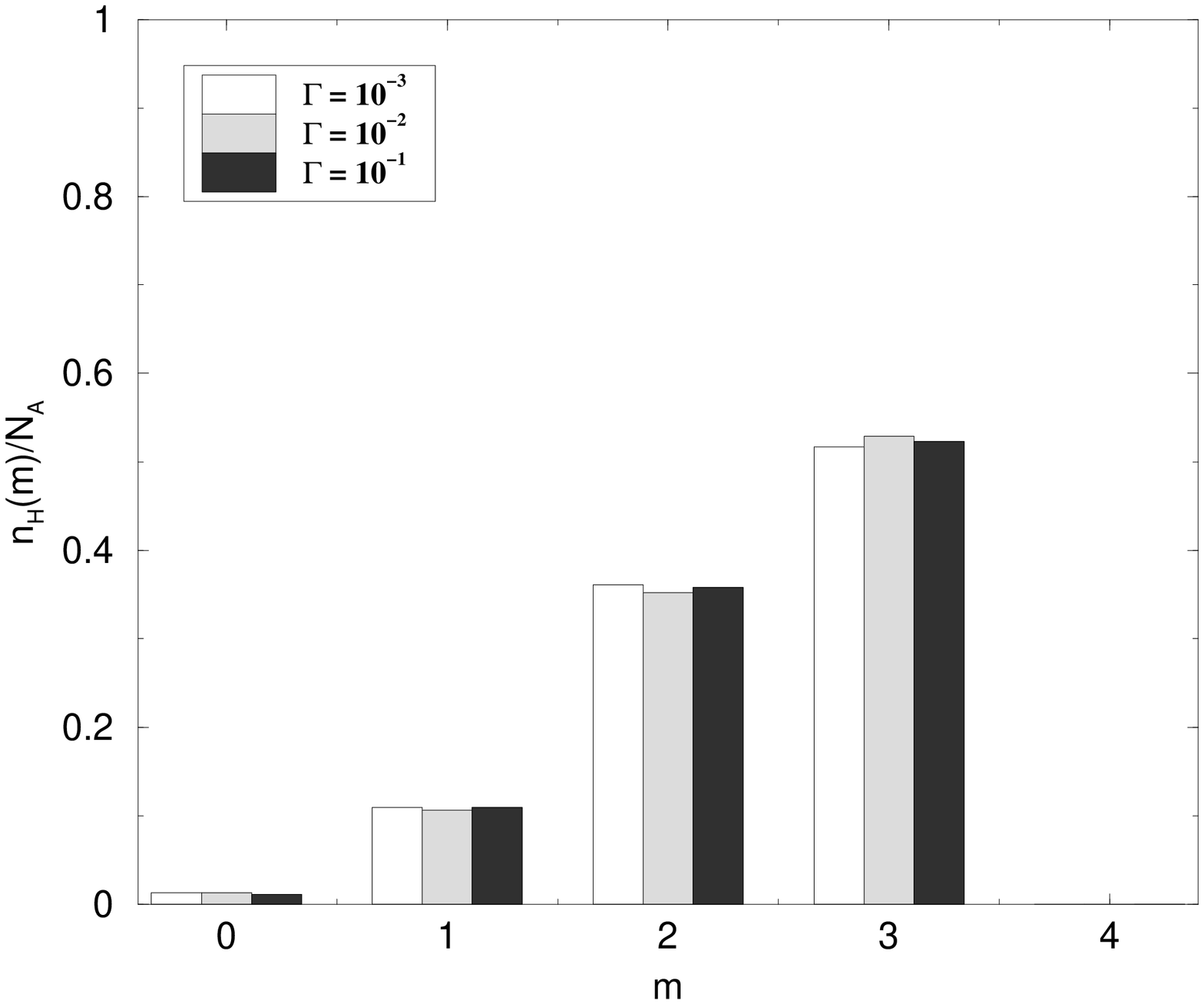}}}
\end{figure} 
\begin{figure}[!ht]
\leavevmode
\rotatebox{270}{\scalebox{.4}{\includegraphics{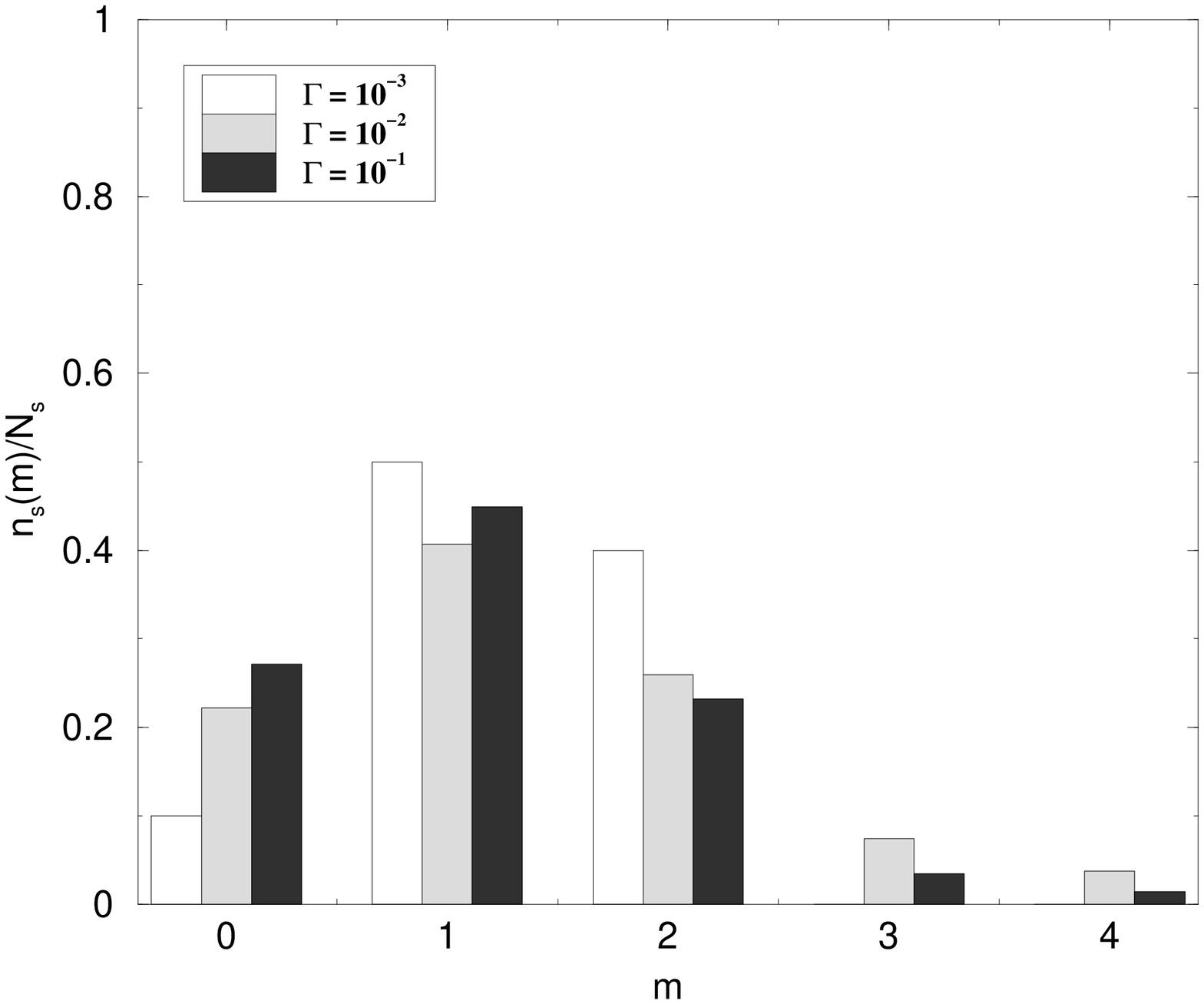}}}
\caption{The distribution of the a) asexual and b) sexual population with 
respect 
to $m$,  for different values of $\Gamma$ for Model $A$ with
hereditary sexuality. The histograms are normalized to unity.}
\end{figure} 

\subsubsection{Mutating the sex gene with constant probability - Model $B$}

Our second strategy for conversion to sex involves a constant  probability 
$\sigma$ for the accidental 
conversion 
to sex, independently  of the distance, as expressed by $m$,  from the 
wildtype.
For this model (Model $B$), once the asexual steady state is reached,
 at each generation we 
allow the sex gene to be ``turned on'' irreversibly, with a small probability 
$\sigma$ for 
each individual. Like in Model $A$,  these individuals will
be ``sexually active'' and  mate with other 
sexually active individuals of that generation.
(If there is only one active sexual at a certain 
time 
step then it has to wait till it finds a partner at a subsequent generation.)
If we take sexual reproduction to be non-hereditary, after mating  the sexual 
individual 
becomes sexually inactive. 
(Within some subsequent generation it can once more become sexually active 
with 
probability
$\sigma$). The deficit in the population is made up by copying randomly 
selected asexual individuals. 


\begin{figure}[!ht]
\leavevmode
\rotatebox{270}{\scalebox{.4}{\includegraphics{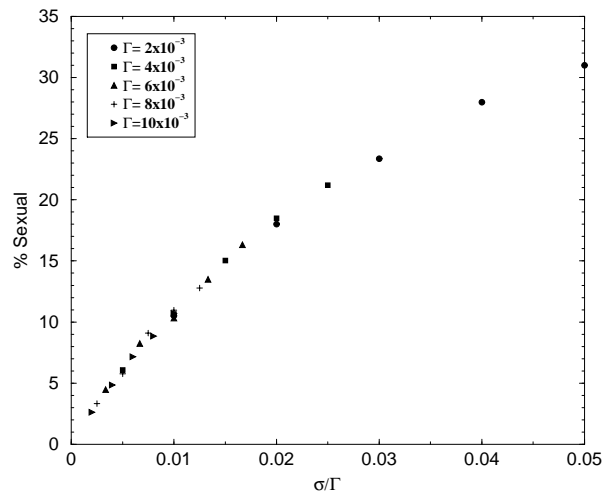}}}
\caption{Percentage of the sexual population v.s. $\sigma / \Gamma$ plots for 
various $\Gamma$
values for Model $B$. Hereditary sexuality is not allowed.  All the 
points collapse onto a single curve in the interval shown.}
\end{figure} 


We find, (see Fig. 5) that this scenario again gives rise to a steady state 
macroscopic 
population of sexuals - but it is smaller than the one in Model $A$. The total 
percentage of  sexuals is a function of $\sigma/\Gamma$, as can be seen from 
the figure, 
and grows with $\sigma/\Gamma$. In Fig. 6(a,b), we display the distribution of 
asexual 
and sexual individuals over the effective number of mutations $m$, for two 
small  values of $\sigma$ 
and $\Gamma$. The characteristic sandpile like~\cite{Jan2} distribution of 
asexuals is 
accompanied by a distribution of sexuals which is again shifted towards 
smaller values of $m$.
It is interesting to observe that
raising $\sigma$ increases the total number of sexuals, and therefore 
depresses the number of asexuals, 
 as is to be expected.  However, it is not immediately obvious why 
keeping $\sigma$ fixed and decreasing the overall mutation rate 
should decrease the number of asexuals. Clearly, raising $\Gamma$ increases 
the 
death rate of both types of organisms, but since the conversion to sex is not 
coupled to 
the increase in the number of mutations, an increased $\Gamma$ only benefits 
the asexuals 
who get cloned to make up the deficit population.
 For large values of $\sigma$,
a novel phase transition takes place, which is the subject of Section IV. 


\begin{figure}[!ht]

\leavevmode
\rotatebox{270}{\scalebox{.4}{\includegraphics{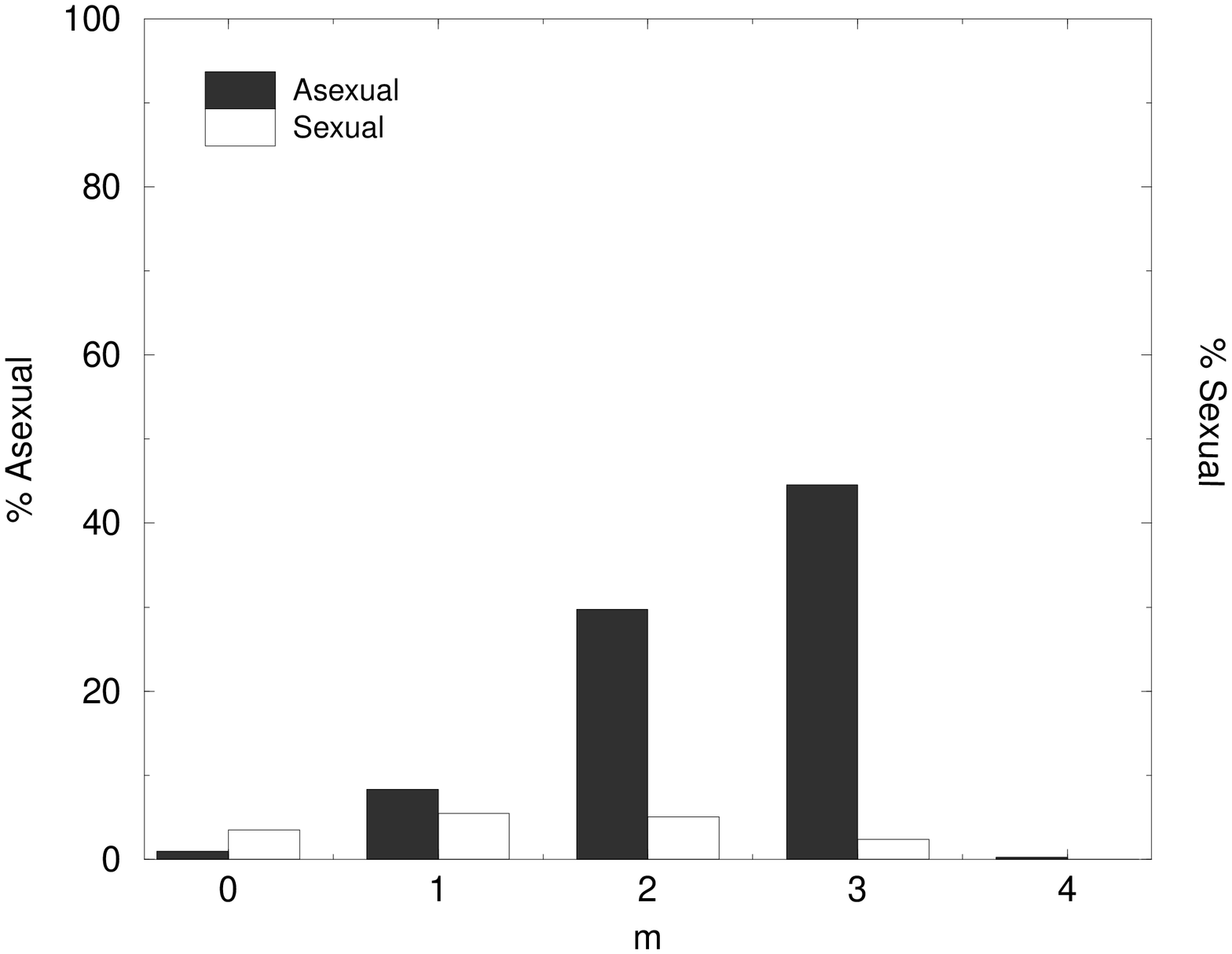}}}
\leavevmode
\rotatebox{270}{\scalebox{.4}{\includegraphics{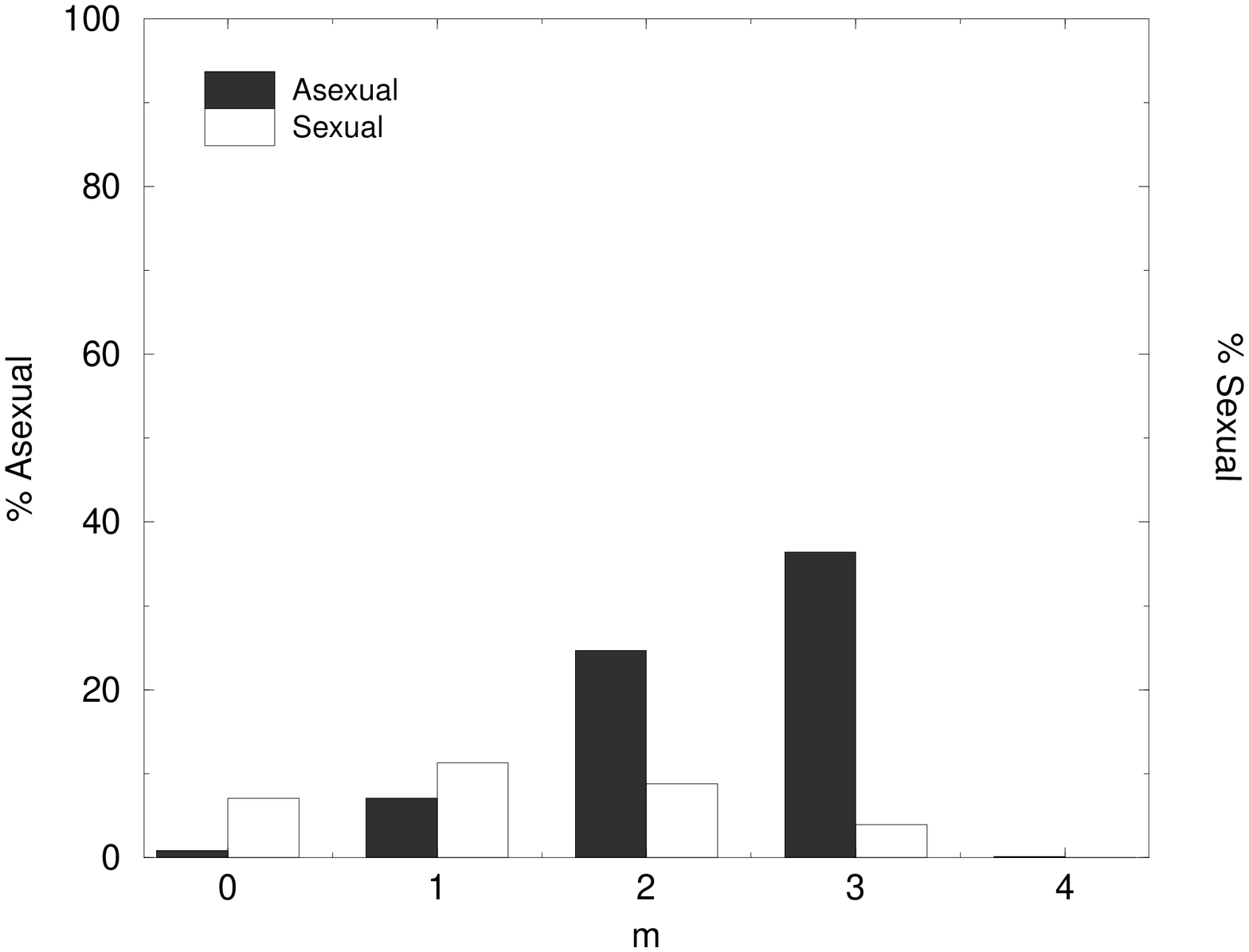}}}
\caption{Distribution with respect to $m$, for both sexual and asexual 
populations 
for Model $B$, without hereditary sex.
a) $\Gamma=6 \times 10^{-3}$, $\sigma = 10 \times 10^{-5}$ and 
b) $\Gamma=2 \times 10^{-3}$,
$\sigma = 10 \times 10^{-5}$;
the histograms represent averages over 10 runs for a population of 1000.}
\end{figure} 


\begin{figure}[!hb]
\leavevmode
\rotatebox{270}{\scalebox{.4}{\includegraphics{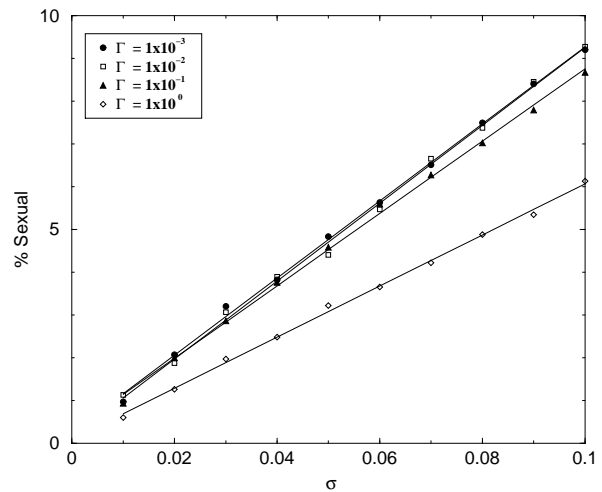}}}
\caption{The percentage of the sexual population v.s. $\sigma$ for various 
$\Gamma$
values for Model $B$ with hereditary sex. The growth with 
$\sigma$ is linear for the different $\Gamma$ values.}
\end{figure} 


{\it Model $B$ with Hereditary Sex}

If the conversion to sexual reproduction is hereditary, then at any given time 
step 
all the sexual individuals mate, except for the odd guy out.
  In Fig. 7 we show the total percentage of the 
sexual population as a function of $\sigma$ alone.  One sees that 
 the growth is very close to linear with $\sigma$, however the collapse as a 
function of $\sigma/\Gamma$  does not occur here.  The curves extrapolate 
to zero at $\sigma=0$. As long as $\sigma>1/N$  one may have a 
small but nonvanishing sexual population.  For smaller values of $\sigma$, the 
number of sexual 
individuals
again fluctuates very strongly and is of $O(1).$ (see Section IV).

\section {Mean Field Evolution Equations} 

To try to understand analytically some of the features found from the 
simulations, 
we have examined the behaviour of the iterative equations that can be obtained 
for the different densities involved. 

Let us define the mutation matrix for haploids, 
$T_{m,m+1}(\Gamma)=\Gamma (L-m)/L$ and 
$T_{m,m-1}(\Gamma)=\Gamma m/L$. Note that $\sum_{\delta=\pm 1} 
T_{m,m+\delta}(\Gamma)= \Gamma$. All the other elements of this matrix are 
zero.  

For low temperatures and for $\mu$, the upper limit of the number of 
mutations tolerated by the haploid individual, being set to four, the survival 
probability is given by,
\be P(m)=\cases{1, m=0,1,\ldots 3\cr
                \yarim, m=4 \cr
                     0, m > 4 \;\;\;. } \label{(P(m)}\ee

\subsection{Asexual steady state}
The time-evolution equations for the asexual population, with 
$n_H(m)$ being the number of individuals with $m$ mutated genes, 
are
\bea
&n_H&(m,t+1) = (1-\Gamma) n_H(m) \cr 
&+&\sum_{\delta=\pm 1} T_{m+\delta,m} n_H(m+\delta,t) - [1-P(m)]n_H(m,t) \cr
&+&\sum_{m^\prime}[1-P(m^\prime)]n_H(m^\prime,t)n_H(m,t)/N_A\;\;\;.\label{asexev}\eea
The last term is the source term, arising from  the 
replacement  of the deceased individuals by randomly cloning the 
extant ones and $N_A = \sum_m n_H(m)$ is the total number of asexual 
individuals.

For large $\beta$, one effectively has,
\bea
n_H(m,t+1) &=& (1- \Gamma) n_H(m,t) \cr 
&+&\sum_{\delta=\pm 1} T_{m+\delta,m} n_H(m+\delta,t) \cr 
&+&\yarim n_H(4,t)n_H(m,t)/N_A\;\;\;,\label{asexev1}\eea
for $ m < 4$. The source term $[1-P(4)] n_H(4)n_H(m)/N_A$ has been replaced by 
its value
$\yarim n_H(4)n_H(m)/N_A$,
and it is 
assumed  that $n_H(m>4) \equiv 0$. This assumption is supported by 
numerical data in the steady state. 

 Note that for  $\Gamma N \sim O(1)$), $n_H(4)$ will 
be  small, i.e.,  of the order of unity.
For $m=4$, this enables us to put the source term in the last 
equation equal to zero, since it will be of $O(1/N)$ while  the 
other terms are of $O(1)$, and we get,
\bea
n_H(4,t+1) &=& (1-\Gamma) n_H(4,t) \cr
&+& \sum_{\delta=\pm 1} T_{4+\delta,4} n_H(4+\delta,t) 
-\yarim n_H(4,t)\cr
&=& (1-\Gamma) n_H(4,t)+\Gamma (1-3/L) n_H(3) \cr 
&-&\yarim n_H(4,t) \;\;\;.\label{asexev4}\eea
Then we see that in the steady state, one may replace 
$n_H(4)/2$ appearing in the source terms by $\Gamma [(1-3/L) 
n_H(3) - n_H(4)]$.  This leads to equations that are homogenous in 
$\Gamma$ in the steady state, yielding, therefore, a steady state 
distribution of the population between  sexual v.s. 
asexual individuals which are independent of $\Gamma $ at least for 
$\Gamma \ge 1/N.$ (see Fig. 1) Iterating these equations leads to a steady 
state with 
an $m$-distribution that is in agreement with the simulation 
results~\cite{Orcal}. 

\subsection{Coexisting asexual and sexual populations}

We now define 
a new quantity, $n_D(m)$ as the number of  $m$-mutation strings  
that {\em belong to a diploid organism}.  The expected number of diploid 
organisms with 
$m$ expressed deleterious mutations can be obtained, once the $n_D(m)$ are 
known.

The probability  for two strings with $m_1$ and $m_2$ mutations (i.e., bit set 
to ``1") to give rise to 
$m$ loci at which both bits are ``1" can easily be calculated. It is given by

\bea
\lefteqn{p(m;m_1,m_2)=} \cr
  & &~~~~~ { m_1! m_2! (L-m_1)! (L-m_2)!\over L! m! (m_1-m)! (m_2-m)! 
(L-m_1-m_2+m)!}
\;\;\;,\nonumber \\
\label{pmmm}
\eea
for $L-m_1-m_2+m > 0$ and $0$ otherwise.
This expression is symmterical in $m_1$ and $m_2$, both of which must be $\ge 
m$.
The number of diploid organisms with $m$ expressed mutations is then, 
\bea
n_s(m)=\yarim \sum_{m_1=m}^{L}\sum_{m_2=m}^{L^*} &p&(m;m_1,m_2)\cr
&\times& n_D(m_1) n_D(m_2)/(2N_S)\;\;\;,\nonumber \\
\eea
where $N_S$ is the ~number of ~diploid ~organisms,  
$\sum_{m=0}^L n_s(m)$, and $L^*= \min[L, L+m-m_1]$.  
The factor of $\yarim$ out front 
comes from converting from the number of gametes that are members of diploid 
organisms with $m$ expressed mutations, to the number of such diploid 
organisms. 
The factor $ n_D(m_2)/(2N_S)$ in the sum 
is the probability of encountering a gamete with $m_2$ mutations as the other 
member of the pair making up the 
diploid organism. 

A similar computation leads to the number of diploid individuals who die as a 
result of too many mutations, 
\bea
 D_D= \yarim \sum_{m=0}^L \sum_{m_1=m}^{L}\sum_{m_2=m}^{L^*}[1-P(m)] 
p(m;m_1,m_2)
\cr
\times~ n_D(m_1) n_D(m_2)/(2N_S)\;\;\;,
\eea
where $L^*$ us defined as above.

The number of gametes with $m$ mutations, which get removed because they 
happen to be members of diploid organisms 
which die, is
\bea
 d_m =\sum_{m^{\prime\prime}=0}^{L} 
\sum_{m^\prime=0}^{\min[m,m^{\prime\prime}]}
[1-P(m^{\prime})] p(m^\prime; m, m^{\prime\prime}) \cr
\times ~n_D(m,t)n_D(m^{\prime\prime}, t)/(2N_S)\;\;\;.
 \label{die-m}\eea

We must also define 
the number of gametes with $m$ bits set to ``1,"
that can take part in sexual reproduction, which is 
\be
\tilde d_m= \sum_{m^\prime}^{\tilde L} p(4;m,m^\prime) 
n_D(m,t)n_D(m^\prime,t)/(2N_S)\ee
where $\tilde L=\min[L, L+4-m]$.  Since  $\tilde d_m$ is only defined for 
$m\ge 4$,  $\tilde L=L+4-m$. 
Note that  
$\sum_{m=4} \tilde d_m= 2 n_s(4)$.

Here we have only considered the scenarios without habitual sex.

{\bf Model $A$}

We now have, from (\ref{asexev1},\ref{asexev4}), for sufficiently large $\beta$
\bea
\lefteqn{n_H(m,t+1) = (1-\Gamma)n_H(m,t)} \cr  
& & ~~+\sum_{\delta=\pm 1} T_{m+\delta,m} n_H(m+\delta,t) - 
\delta_{m,4}n_H(4,t) \cr 
& & ~~+[{3\over 4} n_H(4,t) +D_D(t) + {1\over 4} n_s(4,t) ] 
n_H(m,t)/N_A\;\;.\nonumber \\
\label{asexevA}\eea 
The terms proportional to $\Gamma$ are due to random mutation. The coefficient 
of the Kroenecker 
delta $\delta_{m,4}$ is   $n_H(4)$   since 
all of the asexuals with $m=4$ are removed either due to death or conversion 
to sexuals.
The final term represents the number of $m$-mutation haploids which get cloned 
to keep the population constant;
the expression  in the square brackets is the number of individuals which get 
removed from the population and 
determines  the 
strength of this source term.
The $(3/4)$ factor multiplying $n_H(4)$ comes from two parts: $(1/2)$ of the 
haploids with 4 mutations die;
the other half is converted to sex, and mate,  their number being once more 
halved as a result, contributing $ (1/4)n_H(4)$ 
to the ``removals." $D_D$ (which is $=(1/2) n_s(4)$ for large $\beta$) 
is the number of diploids that die, and $(1/4)n_s(4)$ comes from half of the 
$m=4$ diploid population
being converted to sex, their number being once more halved when they mate.

The dynamics of the number of strands $n_D(m)$ that make up diploid organisms 
is,
\bea
\lefteqn{n_D(m,t+1)=(1-\yarim\Gamma)n_D(m,t)} \hspace{1.5cm}\cr
& & ~+\sum_{\delta=\pm 1} T_{m+\delta,m}(\yarim\Gamma) n_D(m+\delta,t) \cr
& & ~-  d_m(t)-  {1\over 4} \tilde d_m +\delta_{m,4}  P(4) 
n_H(4,t)\;\;\;.\label{sexevA}\eea
For diploids, the probability of a mutation hitting any one gene is halved, 
because there are twice as many
of them. The $d_m$ term is the number of $m$-gametes that are removed as a 
result of death, and in practice 
(for large $\beta$) is nonzero only for $m\ge 4$.
The next term gives the reduction in the number of $m$-gametes as a result of 
sexual reproduction. 
A factor of  (1/2) comes from the probability to engage in sex, and another 
from the fraction of gametes 
that are discarded as a result. Finally, there is a contribution from the 
conversion of haploids to diploids.
We have neglected  the situation where {\it a)}
there is only one active sexual individual is present, so that no mating with 
concomitant 
discarding of a gamete, can take place; or {\it b)}a conversion from haploid 
to diploid 
is impeded because there is only one haploid strand with 4 mutations.  It can 
be checked explicitly that 
Eqs.(\ref{asexevA},\ref{sexevA}) conserve the  total population.

Iterating these equations leads to a steady state distribution that is roughly 
comperable but not identical 
to the simulation results (see Table I).  For $ \Gamma=10^{-3}$ the percentage 
of the sexual population is $24 \%$
of the total, and saturates to $36 \%$ as $\Gamma$ is increased, as compared 
to $70 \%$ from the simulations.
This discrepancy seems to come from the fact that the dynamics is really 
driven by the strongly fluctuating
small population at $m=4$, and mean field theory is simply not able to capture 
this.  

\begin{table}
\caption{
The distribution of the population with respect to the number of 
expressed mutations, obtained from an iteration of the mean field equations
for Model $A$.}
\begin{tabular}{cccccc}
\hline\hline 
 \multicolumn{3}{c}{$\Gamma=10^{-3}$} & \multicolumn{3}{c}{$\Gamma=10^{-2}$} \\
\hline
m     & Asexual\% & Sexual\% & m     & Asexual\% & Sexual\% \\ \colrule
0     & 0.9       & 8.5      & 0     & 0.8       & 9.4  \\
1     & 7.8       & 11.0     & 1     & 6.5       & 16.2 \\
2     & 26.7      & 4.4      & 2     & 22.3      & 8.8  \\
3     & 40.1      & 0.6      & 3     & 33.8      & 1.9  \\
4     & 0.0       & 0.0      & 4     & 0.1       & 0.2  \\ \colrule 
Total & 75.5      & 24.5     & Total & 63.5      & 36.5 \\ \hline\hline
\end{tabular}
\end{table}
The distribution over $m$ is also modified; one sees that the distribution of 
the asexuals is quite similar to 
the simulation results, while the peak of the sexual distribution  has shifted 
to $m=1$, from $m=2$.  This 
indicates that the mean field theory overestimates the effect of remixing, as 
is to be expected, since 
the gametes, instead of being paired in a definite way at any given moment, 
are perpetually part of a 
single gene pool.

{\bf Model $B$}

In this case we have a uniform probability for conversion to sex.  The 
equations become, 
\bea
\lefteqn{n_H(m,t+1) = (1-\Gamma)n_H(m)} \hspace{0.2cm} \cr  
& & ~+\sum_{\delta=\pm 1} T_{m+\delta,m} n_H(m+\delta) - [1-P(m)] n_H(m,t)\cr 
& & ~- \sigma n_H(m) +\{\sum_{m^\prime} [1-P(m^\prime)] n_H(m^\prime) 
+\yarim \sigma N_A \cr
& & ~+ D_D(t) +\yarim \sigma N_S(t)\} n_H(m)/N_A\;\;\;.\label{asexevB}\eea 

Here, haploids are converted to diploids and removed at the rate of $\sigma$, 
and the reduction in the 
population due to mating of recent converts gives the $\yarim \sigma N_A$ term 
in the source. 
The sexuals moreover mate among each other with probability $\sigma$, which 
leads to a further 
sink with strength $\yarim \sigma N_S$.
 Apart from 
these, the terms are  identical to Eq.(\ref{asexevA}).
The dynamics of the $m$-gametes are, 
\bea
\lefteqn{n_D(m,t+1) = (1-\yarim\Gamma)n_D(m,t)} \hspace{1cm}\cr
& & ~+\sum_{\delta=\pm 1} T_{m+\delta,m}(\yarim\Gamma) n_H(m+\delta,t) - d_m
\cr   
& & ~-\yarim \sigma n_D(m,t) + \sigma n_H(m)\;\;\;.\label{sexevB}\eea

In this case, the iteration of mean field equations yield results (see Table 
II)
 that are much closer to those 
found from the simulations.  

\begin{table}
\caption{
The distribution of the population with respect to number of 
expressed mutations, obtained from an iteration of the mean field equations
for Model $B$.}
\begin{tabular}{cccccc} 
\hline\hline
 \multicolumn{3}{c}{$\sigma/ \Gamma=0.01$} & \multicolumn{3}{c}{$\sigma/
\Gamma=1.00$} \\
\hline
m     & Asexual\% & Sexual\% & m     & Asexual\% & Sexual\% \\ \colrule
0     & 2.9       & 9.3      & 0     & 1.7       & 32.2 \\
1     & 14.3      & 3.0      & 1     & 8.3       & 14.7 \\
2     & 32.4      & 0.4      & 2     & 18.8      & 2.3  \\
3     & 37.7      & 0.0      & 3     & 21.9      & 0.1  \\
4     & 0.1       & 0.0      & 4     & 0.0       & 0.0  \\ \colrule 
Total & 87.3      & 12.7     & Total & 50.7      & 49.3 \\ \hline\hline
\end{tabular}
\end{table}
The evolution equations, which we have written as difference equations, are of 
course nonlinear. 
In the simplest case of asexual reproduction (Eqs. (\ref{asexev1}, 
\ref{asexev4})) 
this second order nonlinearity comes purely from the condition of a fixed 
finite population, and appears in the 
source term for the restoration of the population to its fixed value by 
randomly sampling the asexual 
population and cloning it.  With the introduction of sex, 
the source term in the equations for the asexual organisms (\ref{asexevA}, 
\ref{asexevB}) acquires a 
contribution from the number 
of sexual individuals that are removed either through death 
or through sexual reproduction.  Such terms contain nonlinearities up to third 
order.
We expect to find nontrivial behaviour in the limit of large nonlinearities in 
these equations,
and this turns out to be the case, as we explore in the next section.

\section{Limits of strong and extremely weak driving, Chaotic behaviour}

\subsection{The limit of strong driving}

After the discussion of the last section, it is natural to expect that the 
nonlinearities present in the 
problem should drive it to chaotic behaviour when their amplitude is 
sufficiently large.  

We have tested the limit of $\Gamma=1$ and found that for Model $A$ with 
hereditary sex, 
the system becomes unstable.  The total asexual population and sexual 
population display
oscillations with a period of 2 time steps.
The  $m$-distributions also oscillate
for both  populations, with the same period, the amplitude of the oscillations 
being much larger 
for the asexuals.
For such large values of $\Gamma$, 
at each time step a large number of asexuals are driven to large $m$ 
values and are converted to sexuals, they mate, and reduce their expressed 
mutations.  
This leads to a 
macroscopic fluctuation in the number of sexuals, with the halving of the 
mating 
population, which then causes a very 
large number of asexuals to be cloned in turn.  The time average of the sexual 
population 
is depressed slightly below
the saturation value as a result, as can be seen in Fig.(3). 
These oscillations are not observed in the iteration of 
the mean field equations.


\begin{figure}[!ht]
\leavevmode
\rotatebox{270}{\scalebox{.4}{\includegraphics{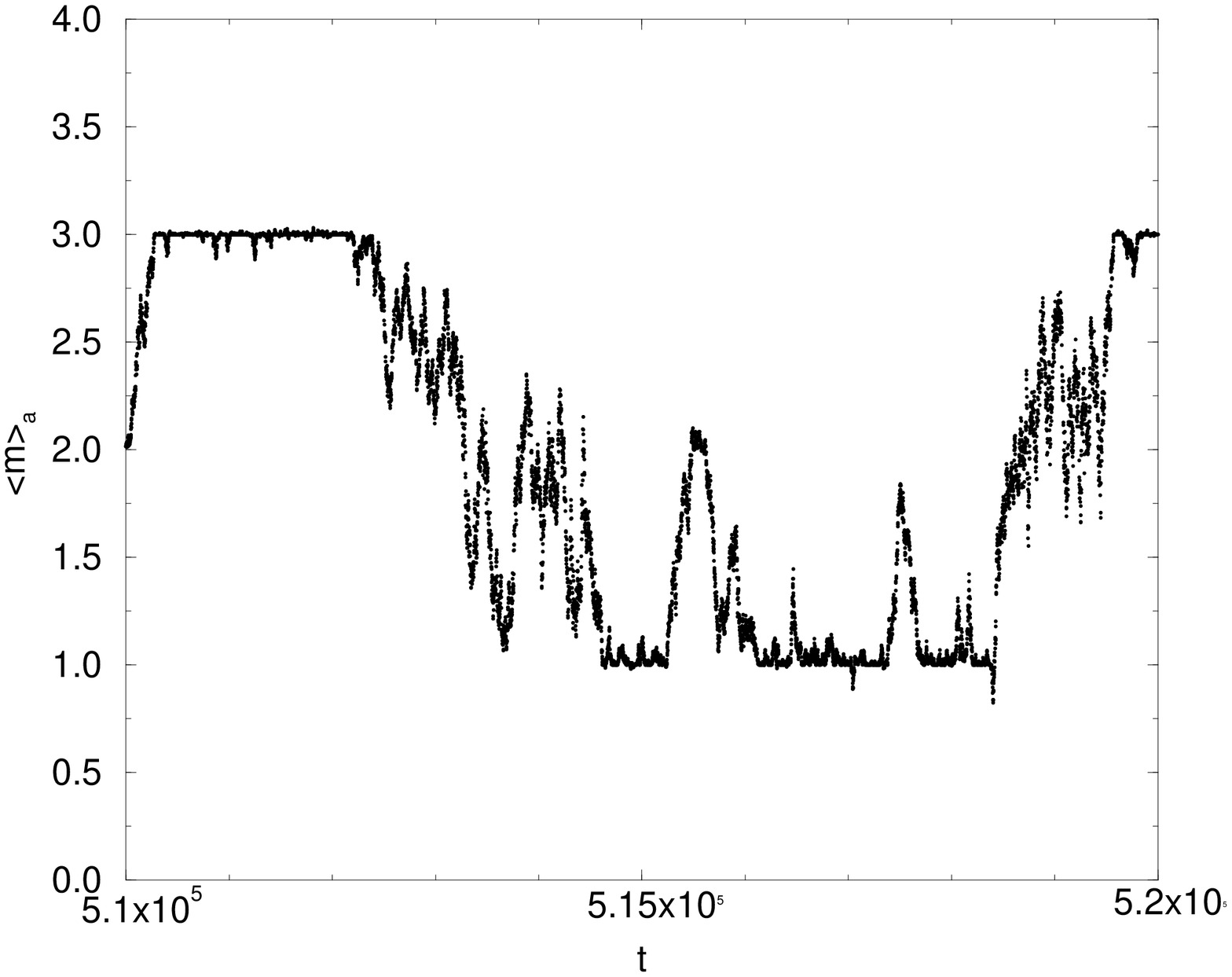}}}
\leavevmode
\rotatebox{270}{\scalebox{.4}{\includegraphics{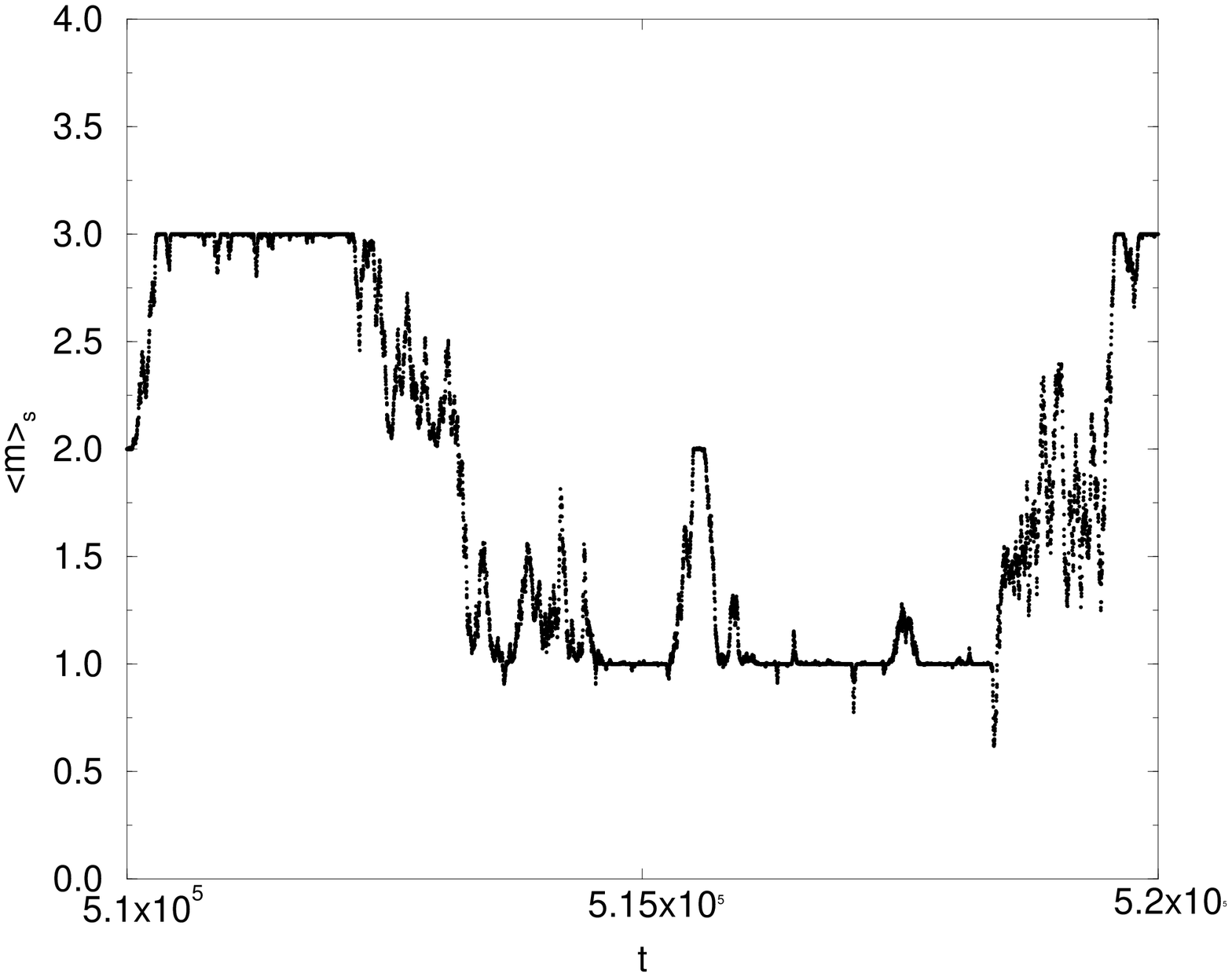}}} 
\caption{In Model $B$, we find intermittent variation with time of a)  
$<m>_a$, the average number of
mutations for the asexual population, and b)  $<m>_s$, the average number of 
expressed mutations
for the sexual population, 
 for the relatively large value of $\sigma=0.5$.  The averages are taken over 
the 
population at time $t$.
$\Gamma=10^{-3}$. It is clearly seen that there are two metastable states.
The pictures show a window of $10^{4}$ time steps
after the transients are dropped.}
\end{figure}


A much more striking behaviour is found in Model $B$ for large values of 
$\sigma$.
As we increase the value of $\sigma$, the probability of random conversion to 
sex, 
 beyond about $0.05$, a spectacular transition takes place to a strange 
attractor for the dynamics of both the asexual and sexual populations. In 
place of the well converged $m$ distributions 
for both asexual and sexual populations, shown in Figs. 6 one observes that 
both distributions are intermittently
switching between several meta-distributions.  The average value of $m$ 
computed over the asexual and 
the sexual populations is shown in Fig.8, and displays this striking 
intermittent behaviour, where 
the distribution of the two populations becomes much more closely coupled than 
in the lower $\sigma$ values. 
They now move more or less {\em in phase,} and their excursions take them all 
the way down to the wild type.  
Now it is only possible to talk about a distribution of distributions.  To 
display this graphically, 
we have plotted the distribution  of  the average 
number of expressed mutations in the two populations, $\langle m\rangle_a$ and 
$\langle m \rangle_s$, as a function 
of $\sigma$.
In Fig.9, we show 
three dimensional plots for these distributions, 
compiled over $10^4$ time steps for each value of $\sigma$. 
In Fig.10, a contour plot of the same distribution as in Fig.9 are shown. 
It is possible to read off from the contour plots that the transition is 
taking place around 
$\sigma_c \simeq 0.05.$


\begin{figure}[!ht]

\leavevmode
\rotatebox{0}{\scalebox{.45}{\includegraphics{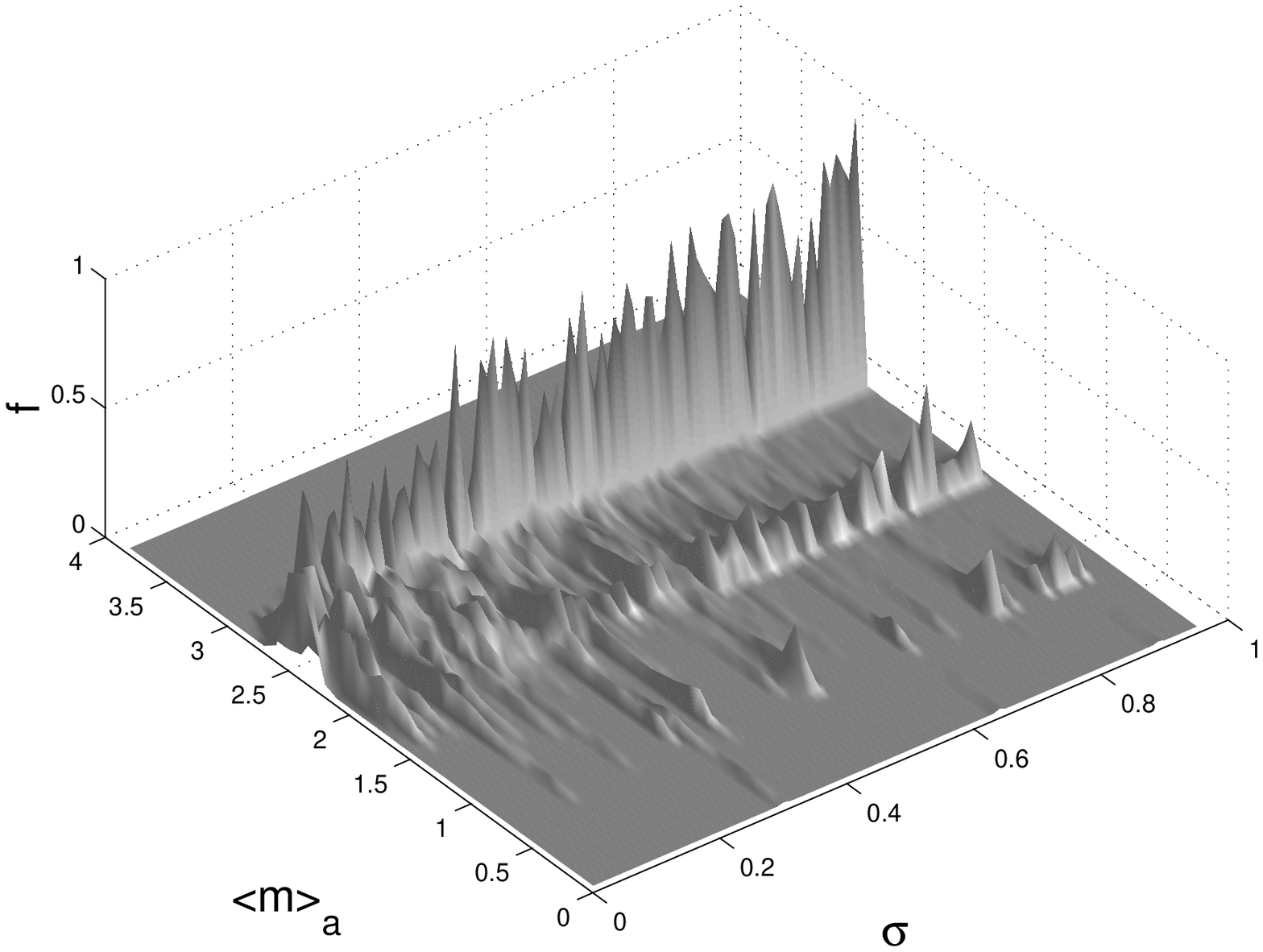}}}
\leavevmode
\rotatebox{0}{\scalebox{.45}{\includegraphics{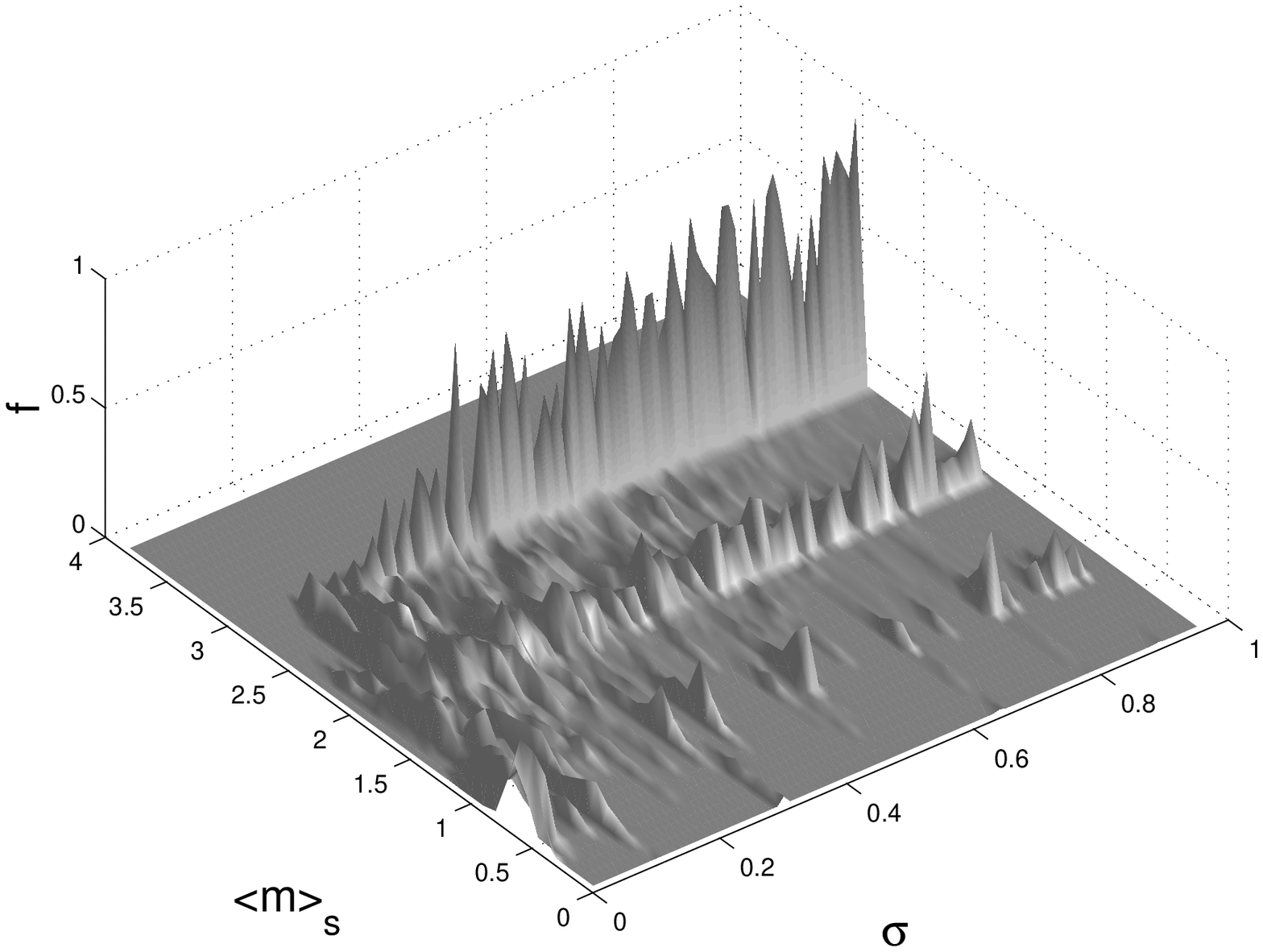}}}
\caption{
A 3D plot showing the branching distributions of a)  $<m>_a$ b)  $<m>_s$
with respect to $\sigma $. After a threshold at $\sigma \sim 0.05$, the
distribution displays more than one peak. The z-axis indicates the relative 
weights of these 
peaks. The total population is 1000 and the figure represents a single 
run of $10^4$ steps.}
\end{figure} 



\begin{figure}[!ht]

\leavevmode
\rotatebox{0}{\scalebox{.4}{\includegraphics{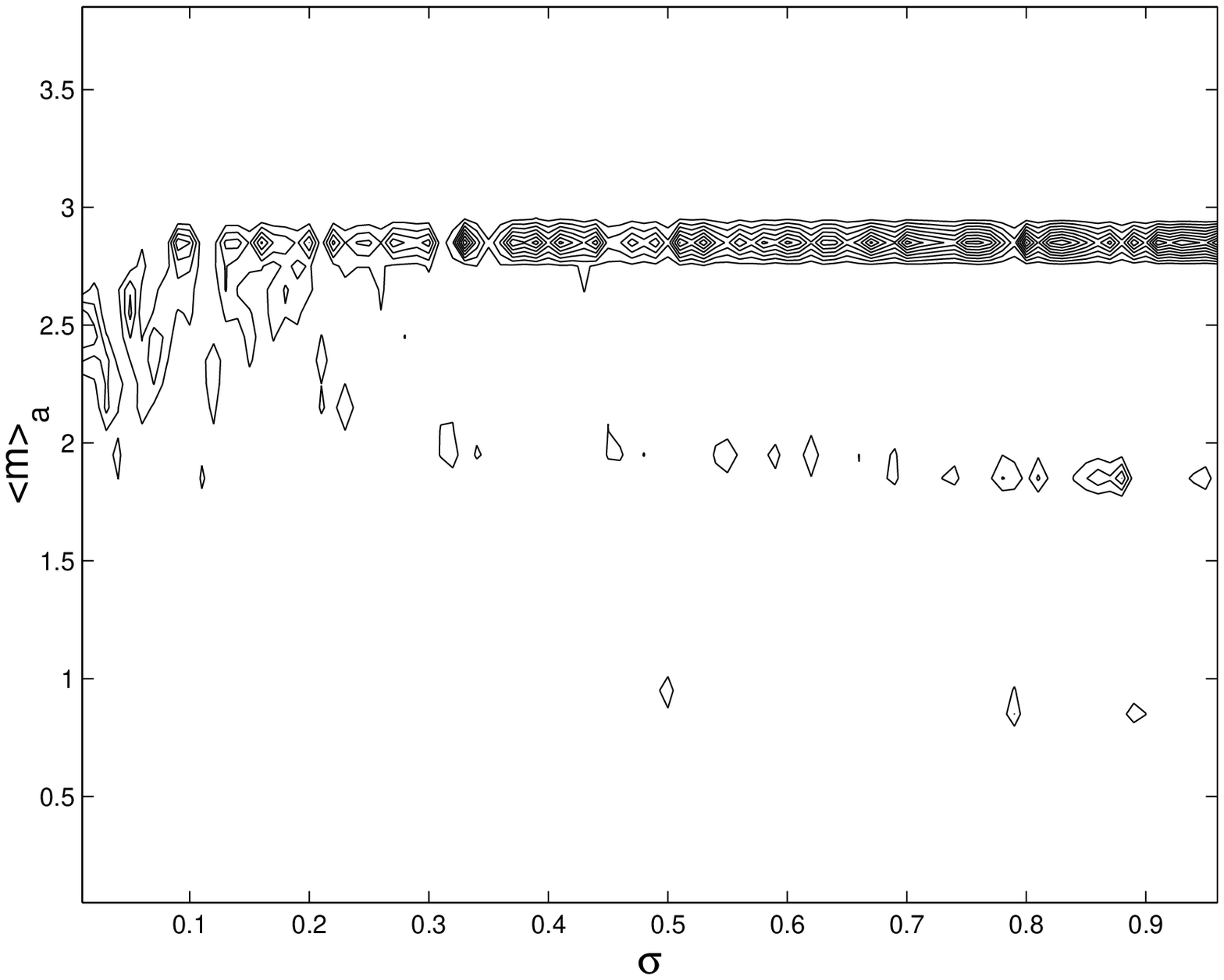}}}
\leavevmode
\rotatebox{0}{\scalebox{.4}{\includegraphics{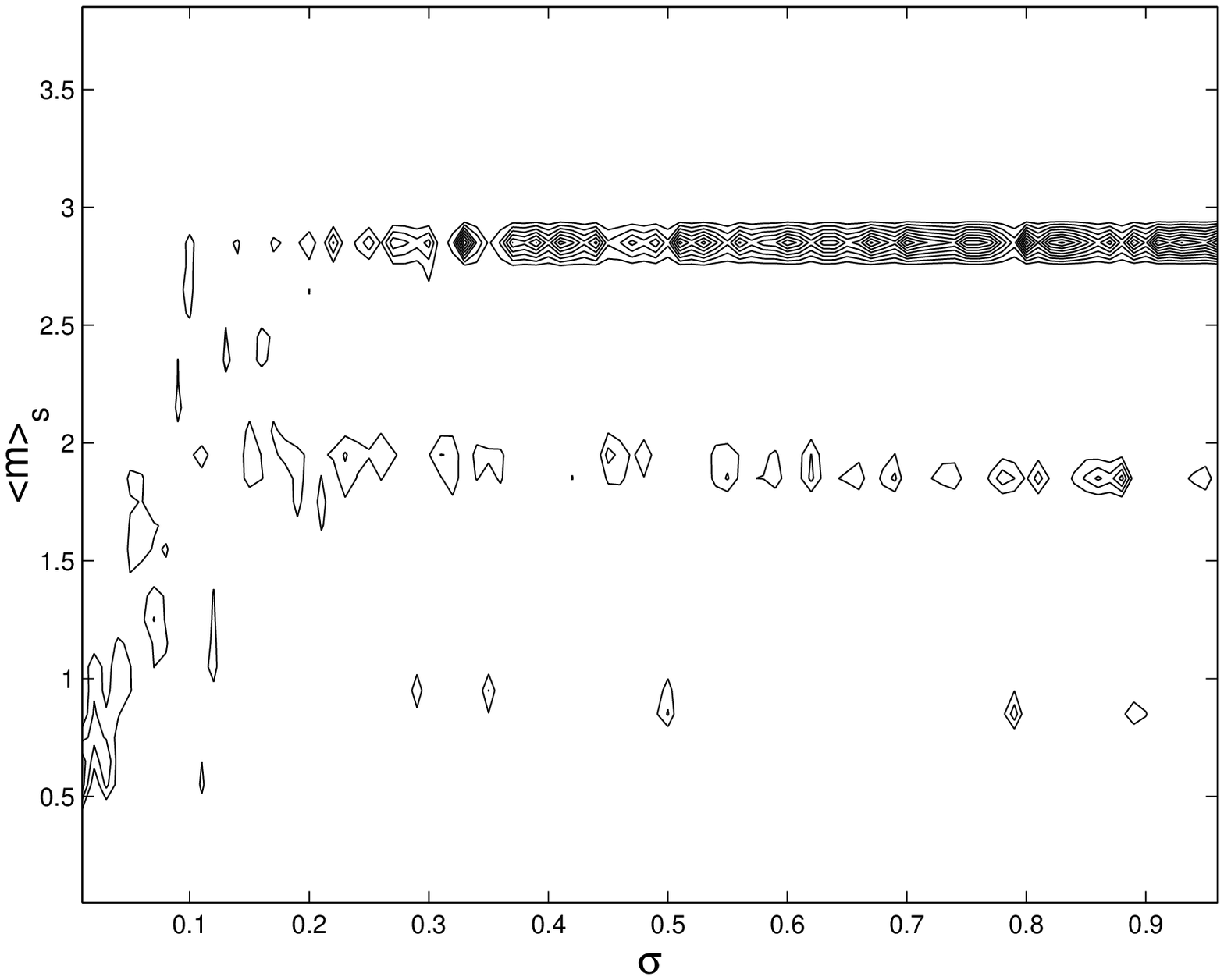}}}
\caption{Contour plot showing the branching of a) $<m>_a$  b)  $<m>_s$ , 
as $\sigma $ increases,
for a population of 1000, computed over $10^4$ time steps.}
\end{figure} 


Besides being intermittent, this transition has a dramatic effect on the $m$ 
distribution of the sexual 
population, in that it shifts it to much higher values.  It can be seen in 
Fig.10(b)
that for $\sigma <\sigma_c$, the mean $m$ for the sexual population is 
$\langle m \rangle_s \sim 0.75$, while for large $\sigma$ it is comparable 
to the corresponding value for the asexual population, closer to 3.  The 
reason seems to be that 
with the great depletion of the population when too many individuals are being 
switched on to sex 
and engaging in sexual reproduction, the asexuals are cloning too many 
identical copies to make up 
for the deficit. When these are subsequently turned sexual and mate among each 
other, ``inbreeding" 
takes place - there is not sufficient genetic diversity for sex to lead to 
sufficient mixing and 
therefore an amelioration of the effective fitness.

We have iterated the mean field equations (\ref{asexevB},\ref{sexevB}) for 
Model $B$ and found that this 
intermittent behaviour is suppressed. The sexuals simply evolve along the 
lower branch which in the simulations
has  the smaller weight, while the asexuals evolve along the higher (large 
$m$) branch,
which has the greater weight in the simulations,
and the evolution is completely stable.  For $\sigma=0.9$ and $\Gamma=0.1$,
 $\langle m \rangle_a=2.43$ and 
$\langle m \rangle_s=0.47$. 

\subsection{The limit of infinitely slow driving ($\Gamma \to 0$ or $\sigma 
\to 0$)}

In the limit of infinitely slow driving, i.e., $\Gamma \to 0$ or $\sigma \to 
0$, we observe a 
transition to a different phase.  


\begin{figure}[!ht]
\leavevmode
\rotatebox{270}{\scalebox{.4}{\includegraphics{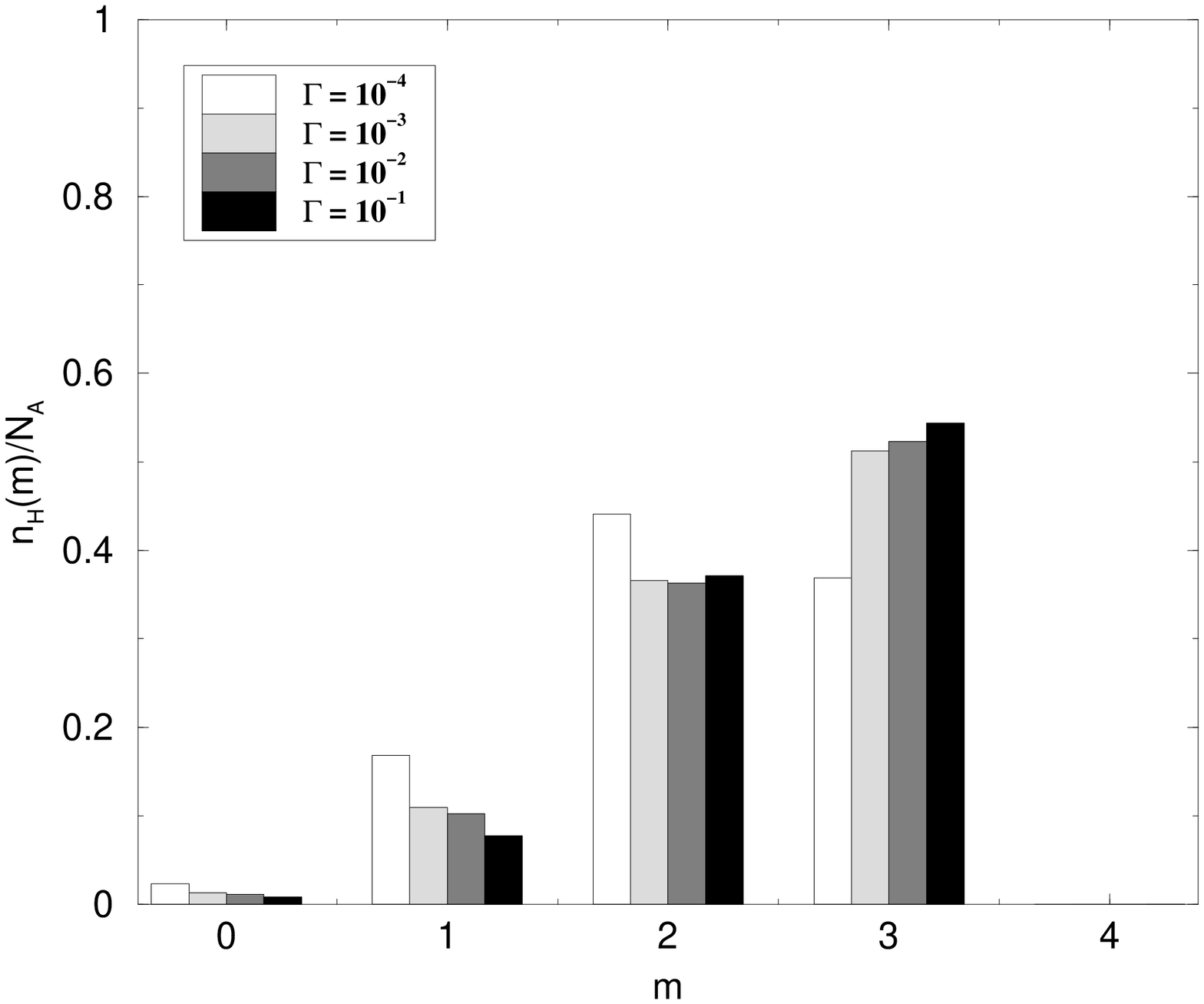}}}
\leavevmode
\rotatebox{270}{\scalebox{.4}{\includegraphics{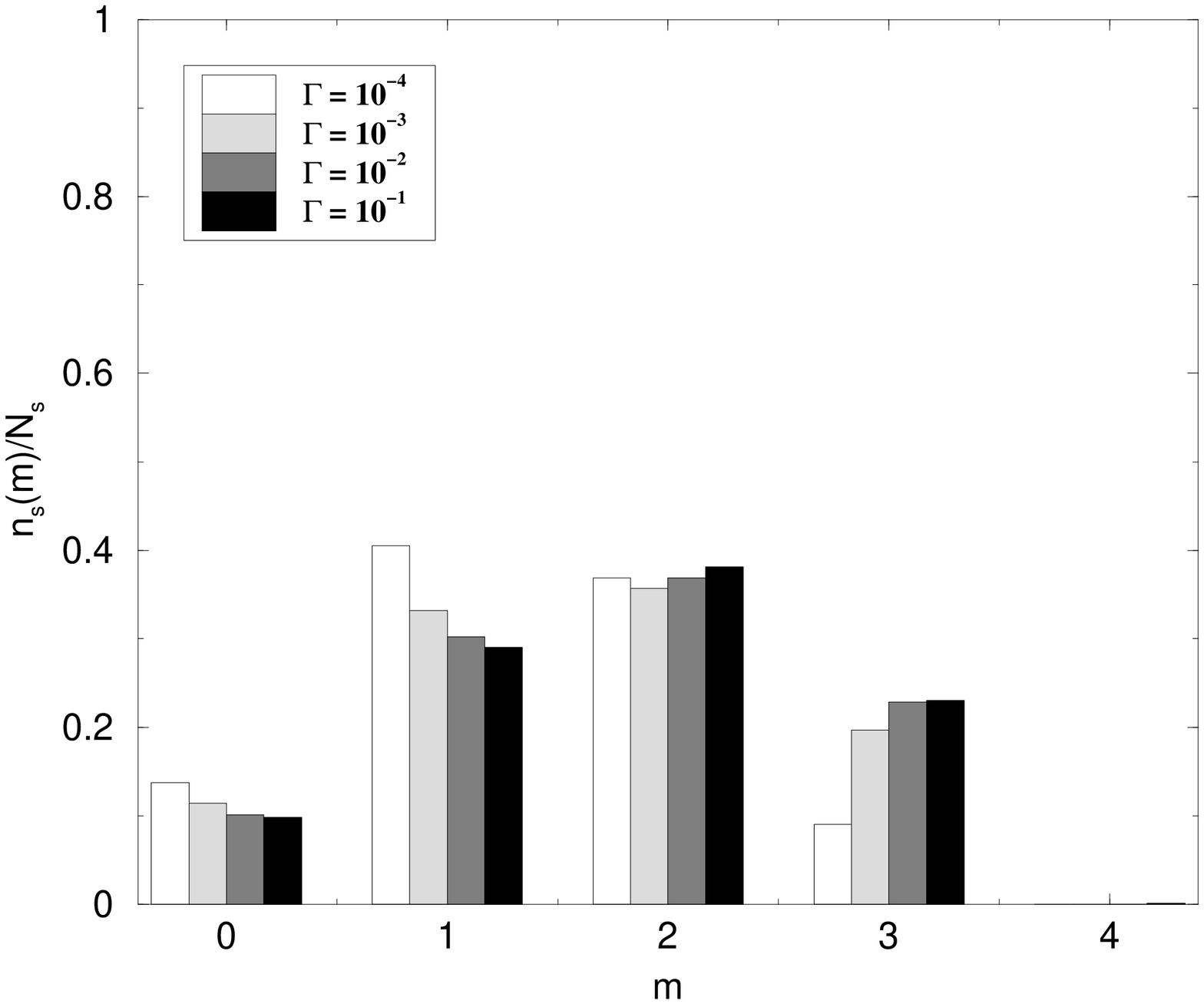}}}
\caption{The distribution over $m$ for a) asexual b) sexual populations, 
for different values of $\Gamma$ for Model $A$, without
hereditary sex. The steady state distribution changes and
the peak on the distribution shifts to a smaller $m$ value as one lowers the
$\Gamma$ value below the threshold $1/N= 10^{-3}$.}
\end{figure}


For $\Gamma < 1/N$, we find a qualitatively 
different asexual steady state, where the $m$ distribution has shifted to 
lower $m$ values (compare 
with Fig. 2 of \cite{Orcal}) and no longer has the characteristic minimally 
stable sand-pile 
like~\cite{Jan2} distribution.  For $\Gamma=10^{-4}=(10 N)^{-1}$, over a run 
of $10^{6}$ steps, 
we find $n_H(m)/N \simeq 0.03, 0.14, 0.44, 0.39$ for $m=0,\ldots,3$ 
respectively, where the peak has moved
to $m=2$ from $m=3$, or broadened towards the left.  This does not seem simply 
to be due to a slowing down of 
the dynamics.  Rather, once the mutation rate drops below $1/N$, the flow over 
the 
$m=4$ threshold which stabilizes the skewed distribution slows down to a 
dribble.  This gives the 
$m$ distribution time  to  get stabilized at $m=2$ rather than being pushed to 
the $m=3$ limit. The mechanism for the stabilization is provided by the dead 
bacteria being replenished
from among the most prevelant extant ones.

Once sex is turned on in Model $A$, we similarly observe that the peaks in the 
distribution of the asexual 
and sexual populations have shifted to lower $m$ values ($m=1$ and $m=2$ 
respectively), as shown in Fig.11.
Although the total sexual population is relatively small here, we have checked 
that the fluctuations in the 
histogram over 10 different realizations stay small.

Iteration of the dynamical equations, on the other hand, reveal no such phase 
transition 
and, for the asexual steady state, converge 
to the same steady state distributions as found for $\Gamma > 1/N$. 
 In Fig. 12, we show the 
time series for $n_H(m)$ ($N=100$)
for the asexual population without conversion to sex.  
At time $t=0$, the largest density is of course 
at $m=0$, and then the maximum shifts successively to $m=1,2$ and finally to 
$m=3$ 
where it stabilizes.  Comparison 
with the simulation results seem to indicate that the simulations get stuck at 
an intermediate 
``metastable" state, while the peak is around $m=2$.  The fact that in the 
simulation 
one has to wait around 
until, with a very low probability, a discrete individual is pushed over the 
$m=4$ 
barrier, dies, and is cloned 
from among the live bacteria, while in the mean-field equations, there is a 
weak but steady seepage, which
prevents this phase transition from taking place.


\begin{figure}[!hb]
\leavevmode
\rotatebox{270}{\scalebox{.4}{\includegraphics{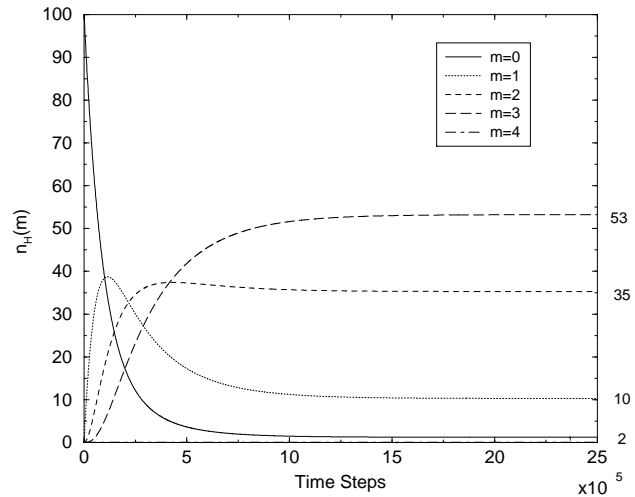}}}
\caption{The iterated solutions of the equations for the purely asexual
population, without the introduction of sex,
as a function of time for different values of $m$. $\Gamma=10^{-4}$. }
\end{figure}

\section{Discussion}

The mechanism of random conversion to sex, in the presence of a constant rate 
of mutations, as 
 investigated in this paper as scenario for the maintanence of a macroscopic 
sexual production, is in fact 
very closely related to ``coevolution of cell senescence and diploid sexual 
reproduction in unicellular organisms," studied by Cui et al.~\cite{Cui}.  
In this 
paper a ``senescence clock" ticks off a finite lifetime for each bit-string.  
Sexual reproduction 
(conjugation) resets the senescence clock; unless this happens after a number 
of generations of cloning, 
the offspring stop dividing and die.

Our Model $B$ can be seen as a simpler version of the model proposed by Cui et 
al., with an intrinsic mechanism, 
provided by Muller's ratchet~\cite{Muller}, for cell senescence. The constant 
mutation rate sets the time 
scale for the survival of 
any given bitstring, unless it succeeds  engaging in sex, with a given 
probability (our $\sigma$). 
A survival function (Eq.(1)) leads to the 
elimination of genomes carried by haploid individuals 
multiplying by asexual reproduction, once they have  
accumulated too many mutations as a result of prolonged exposure to the 
constant mutation rate~\cite{Smith,Muller}.

Our Model $A$ goes one step further, in that it makes the number of mutations 
(the cell clock) provide the 
triggering mechanism for the transition to diploidy and sex.  It is gratifying 
to find that this is a more 
successful strategy for establishing a sexual population than a constant rate 
of conversion to sex.

 Chopard et al.~\cite{Droz} have pointed out that 
 care must be taken in the investigation of finite populations, 
amplifiying and stabilizing small fluctuations which 
in the thermodynamic limit would be attenuated to zero. They emphasize the 
importance 
of spatial variations  
which cannot be captured by mean field theories. 
In this paper we have demonstrated the relevance for finite populations of 
discrete stochastic events, 
whose effect in the very
weak driving limit cannot be captured by the ``mean field" equations.
In the very weak driving limit the system is below the hydrodynamic regime, 
and exhibits a qualitatively different phase than which is described by the 
continuum approximations.

In a recent article Pekalski~\cite{Pekalski} has studied a model which is in 
many ways similar to ours.  
There the success of 
sexual reproduction, meiotic parthenogenesis and asexual reproduction, in 
maintaining a finite population in the 
face of periodically changing environmental conditions and a constant mutation 
rate, is studied in terms of 
the relative sizes of the populations.
Age is included in the model as a parameter which reduces the fitness.  The 
populations do not interact. 
The findings are that meiotic parthenogenesis and sexual reproduction are more 
favorable than mitotic reproduction,
with slight differences between them depending on the precise conditions.

Further work is in progress, to investigate the effect of finite temperature, 
and of including
the possibility of genetic crossover and meiotic parthenogenesis, in our 
models.  Results on the 
autonomous viability of the sexual population, after the steady conversion 
from the 
haploid population has been switched off (but mitosis allowed for the 
diploids), will be 
reported in a future publication.

{\bf Acknowledgements}

It is a great pleasure to thank Naeem Jan, who has introduced us to this 
subject, Dietrich Stauffer for his continuing interest,
Asli Tolun and Candan Tamerler Behar who have helped us battle our ignorance 
of biology. We thank 
 P\i nar 
\"Onder for useful discussions.  
One of us (A.E.) gratefully acknowledges partial 
support from the Turkish Academy of Sciences.

\end{document}